\documentclass[journal, draftcls, onecolumn, 10pt]{IEEEtran}

\ifCLASSINFOpdf
\else
   \usepackage[dvips]{graphicx}
\fi
\usepackage{url}

\usepackage{graphicx}
\usepackage{optidef}
\usepackage{algorithm}
\usepackage{algorithmic}
\usepackage{amsmath,amsfonts,amssymb,amsthm, bm}
\usepackage{mathtools, enumitem}
\usepackage{tikz}
\usepackage{subcaption, float, comment} 
\usepackage{float}
\usetikzlibrary{arrows,decorations.markings, graphs}
\usetikzlibrary{positioning}
\newcommand\norm[1]{\left\lVert#1\right\rVert}
\DeclareMathAlphabet{\mathpzc}{OT1}{pzc}{m}{it}

\DeclareMathOperator*{\argmin}{arg\,min}
\renewcommand{\vec}[1]{{#1}}
\newcommand{\G}{\mathcal{G}}
\newcommand{\D}{\mathcal{D}}

\newcommand{\V}{\mathcal{V}}
\newcommand{\Du}{\mathfrak{d}}
\newcommand{\Tr}{\textsf{T}}
\newcommand{\Tra}{\text{tr}}
\newcommand{\He}{\textsf{H}}

\floatname{algorithm}{Algorithm}
\begin{document}

\title{Investigating the relationship between graph eigenvector ordering and the signal processing dual}

\author{B Subbareddy, S Sai Ashish, Aditya Siripuram 
\thanks{The authors are at the Indian Institute of Technology Hyderabad}}

%\markboth{Journal of \LaTeX\ Class Files, Vol. 14, No. 8, August 2015}
\maketitle

\begin{abstract}
Graph signal processing uses the graph eigenvector basis to analyze signals. However, these graph eigenvectors are typically linearly ordered (by total variation), which may not be reasonable for many graph structures. There have been structure-based similarity metrics proposed in the literature that better capture the geometry of graph eigenvectors. On the other hand, there has been worked that attempts to generalize the concept of duality to graph signal processing. The (signal processing) dual graph captures the relationship between graph frequencies and is obtained by typically inverting the graph Fourier transform operation. In this work, we investigate the connections between these two concepts.

We propose a dualness measure of two graphs, which quantifies how close the graphs are being (signal processing) duals of each other. We show that this definition satisfies some desirable properties and develop an algorithm based on coordinate descent and perfect matching to compute an approximation to dualness. By computing the dualness measure, we observe that for structured graphs, the similarity metric-based techniques give a better dual graph (and vice-versa for ER graphs), suggesting a potentially novel approach unifying the two methods.
\end{abstract}


\IEEEpeerreviewmaketitle

\section{Introduction}

\IEEEPARstart{G}{raph} signal is a real-valued function on the vertex set of a graph. The topology of this graph encodes the inherent ordering or relationship between the signal indices. Such models are suitable for processing data that is non-linearly structured and complex, which is ubiquitous in many applications \cite{ortega2018graph}, including the data from neuro-imaging \cite{bullmore2009complex}, sensor networks, gene networks \cite{ratnasamy1999inference} and transport networks  \cite{shuman2013emerging} etc. Graph signal processing (GSP) \cite{ortega2018graph} has emerged as a generalized framework to extend concepts and techniques from traditional signal processing to graph signals. Similar to how the classical Fourier transform expresses a signal in terms of complex exponentials \cite{osgood2019lectures}, the Graph Fourier transform (GFT) expresses a graph signal in terms of eigenvectors of the adjacency or the Laplacian matrix of the graph. The GFT of a graph signal gives us the graph frequency components, much like the Fourier transform (FT) gives us the frequency components. %Recent work in this field has extended concepts like filtering and sampling to graph signals (see \cite{teke2016extending}, \cite{tay2015techniques} \cite{chen2015discrete}, \cite{anis2016efficient} for a few references).

%Graph signal processing (GSP) \cite{ortega2018graph} has emerged as a principled framework to extend concepts and techniques from traditional signal processing to signals over the graph. GSP handles such complex data by effectively capturing the underlying relationship using graphs. In GSP, signals reside on the vertices of a graph, and the topology of the graph encodes the inherent ordering or relationship between signal components. This encoding inherently means that the signal components may not be linearly ordered; which makes GSP suitable for processing data which is unstructured, complex, and massive. Such data is ubiquitous in many applications \cite{ortega2018graph} including the data from brain networks \cite{bullmore2009complex}, sensor networks, gene networks \cite{ratnasamy1999inference} and transport networks  \cite{shuman2013emerging} etc. 

%Graph Fourier transform (GFT) expresses a graph signal in terms of eigenvectors of the adjacency or the Laplacian matrix of the graph.This transform is similar to classical Fourier transform that expresses a signal in terms of complex exponentials \cite{osgood2019lectures}. The GFT of a graph signal gives us the graph frequency components, much like the Fourier transform gives us the frequency components. The Laplacian based GFT has a frequency or variation based interpretation like the classical Fourier transform \cite{huang2018graph} \cite{shuman2013emerging}.

%\subsection{Ordering of the eigen vectors motivation}

For traditional (linearly ordered) signals, the underlying graph can be represented as a path (or a cycle in the case of circular indices). In this case, the GFT is the same as the classical Fourier transform. From this perspective, the GFT (when applied to arbitrary graphs) can be thought of as a generalization of the Fourier transform. Indeed, a lot of the development in GSP is based on generalizing concepts of filtering \cite{DSPGraphs} \cite{GLFilteredGraphSignals}, denoising \cite{shuman2013emerging}, and sampling \cite{Anis_2016} from traditional signal processing. However, many of the ideas from traditional signal processing, including convolution and domain symmetry (or duality), do not seem to carry over to GSP. In traditional signal processing, the concept of duality between time and frequency domains is an important property of Fourier transforms and has been heavily used in discrete-time analysis \cite{oppenheim1975digital}. Duality leads to the concept of Fourier transform pairs and helps in applying algorithms developed for the time domain to the frequency domain. 

By contrast, in GSP, even though the signal indices are assumed to be non-linearly ordered (i.e., described a graph), the graph frequencies (or eigenvectors) are usually assumed to be linearly ordered \cite{shuman2013emerging} \cite{ortega2018graph}, \cite{sandryhaila2014discrete}: this has consequences for defining frequency bands, smoothness, sampling patterns, and filtering schemes. However, a linear ordering might not best illustrate the underlying relationship between graph frequencies. Consider, for example, a grid graph as in Fig \ref{fig:Ordering_Grid}: the graph eigenvectors will have variation both in the horizontal and vertical directions. Thus a linear (or one-dimensional ordering) does not fully capture the graph frequency space \cite{saito2018can}. Moreover, the existence of sparse eigenvectors for certain graphs (for e.g. star graphs as in Fig \ref{fig:seq_orderL}) \cite{teke2017uncertainty} \cite{chung1997spectral}  implies that the frequency components (especially those corresponding to these sparse eigenvectors) correspond to signals in different regions of the graph, suggesting a potential non-linear ordering of graph frequencies. 

In GSP, ordering the graph frequencies is essential to shape the vertex-frequency symmetry. In particular, we can model the graph frequencies as residing on the vertices of a (frequency domain) graph; to mirror the similar model in the vertex domain. The investigation of such models seems to have taken place in two separate branches in the literature.
\begin{enumerate}
    \item On the one hand, there has been worked on developing distance metrics between graph eigenvectors (using heat diffusion models \cite{cloninger2021dual}, and Ramified Optimal Transport Theory \cite{saito2018can}). This distance metric is then embedded in a low dimensional space using standard techniques to obtain the ordering of the graph frequencies. Note that the distance metric so obtained inherently defines a frequency domain graph.
    \item On the other hand, there has been worked on generalizing the concept of duality from traditional signal processing to GSP \cite{leus2021dual}. We say that graphs $\G_1$ and $\G_2$ are (signal processing) duals - \emph{ideally}, for signals residing on the graph $\G_1$; their GFT resides on the dual graph $\G_2$, and applying GFT again gives us back the original signal on $\G_1$. These techniques obtain the dual graph $\G_2$ by inverting the GFT from $\G_1$. This technique has applications in filter design \cite{leus2021dual}. Note that the obtained frequency domain graph inherently defines an ordering of the graph frequencies.
\end{enumerate}
 Both these approaches ultimately lead to a graph and ordered frequency indices. Thus, it is natural to wonder if the frequency domain graphs obtained by these two (seemingly disparate) branches of investigation are related. In this work, we systematically investigate this relationship and compare the frequency domain graphs obtained by these techniques.

Towards this end, we define the dualness of a pair of graphs $\G_1$ and $\G_2$: the dualness measures how close $\G_1$ and $\G_2$ are to being (signal processing) duals. We propose the dualness measure also provides a systematic framework to analyze algorithms that find a frequency domain graph. For instance, consider the five node graph $\G_1$ in Figure \ref{fig:graphs-dual}. In \cite{leus2021dual}, the authors proposed an algorithm to find a dual, which results in the graph $\G_2$ as in Figure \ref{fig:graphs-dual}). However, the graph $\G_3$ (Figure \ref{fig:graphs-dual}) has a higher dualness measure ($1.728$ as opposed to $0.454$ for $\G_2$), and so is a better candidate for the dual of $\G_1$. Computing such a measure has challenges, e.g., even if the graph vertices are labeled in a different order, The dualness measure is expected to remain the same. We highlight the key challenges and our approach towards overcoming them in Section \ref{sec:dualness_measure}. We then compute the dualness measures for the dual graphs obtained by both the metric-based and signal processing-based techniques. We observe that for structured graphs, the metric-based techniques give a better dualness score, whereas, for ER graphs, the signal processing-based techniques give a better dualness score (Section \ref{sec:Analysis}). This suggests that there may be scope to combine the metric-based and signal processing-based approaches to identify the graph on the frequency indices.

\begin{figure}[h]
\centering 
\begin{subfigure}[b]{0.4\textwidth}
\centering
\begin{tikzpicture}[scale=0.65]
    \node[shape=circle,draw=black, inner sep=1pt, outer sep = 0pt] (3) at (0,0) {3};
    \node[shape=circle,draw=red,draw=black, inner sep=1pt, outer sep = 0pt] (1) at (0,1.5) {1};
    \node[shape=circle,draw=red,draw=black, inner sep=1pt, outer sep = 0pt] (4) at (1.5,0) {4};
    \node[shape=circle,draw=black, inner sep=1pt, outer sep = 0pt] (2) at (1.5,1.5) {2};
     \node[shape=circle,draw=red,draw=black, inner sep=1pt, outer sep = 0pt] (5) at (3,1.5) {5};
    \node[shape=circle,draw=black, inner sep=1pt, outer sep = 0pt] (6) at (3,0) {6};
    \node[shape=circle,draw=black, inner sep=1pt, outer sep = 0pt] (8) at (1.5,-1.5) {8};
    \node[shape=circle,draw=red,draw=black, inner sep=1pt, outer sep = 0pt] (9) at (3,-1.5) {9};
    \node[shape=circle,draw=red,draw=black, inner sep=1pt, outer sep = 0pt] (7) at (0,-1.5) {7};
    \node[shape=circle,draw=red,draw=black, inner sep=1pt, outer sep = 0pt] (10) at (4.5,-1.5) {10};
    \node[shape=circle,draw=red,draw=black, inner sep=1pt, outer sep = 0pt] (11) at (6,-1.5) {11};
    \node[shape=circle,draw=red,draw=black, inner sep=1pt, outer sep = 0pt] (12) at (7.5,-1.5) {12};
    \node[shape=circle,draw=black, inner sep=1pt, outer sep = 0pt] (13) at (4.5,0) {13};
    \node[shape=circle,draw=black, inner sep=1pt, outer sep = 0pt] (14) at (6,0) {14};
    \node[shape=circle,draw=black, inner sep=1pt, outer sep = 0pt] (15) at (7.5,0) {15};
    \node[shape=circle,draw=red,draw=black, inner sep=1pt, outer sep = 0pt] (16) at (4.5,1.5) {16};
    \node[shape=circle,draw=red,draw=black, inner sep=1pt, outer sep = 0pt] (17) at (6,1.5) {17};
    \node[shape=circle,draw=red,draw=black, inner sep=1pt, outer sep = 0pt] (18) at (7.5,1.5) {18};

    \path [-] (3) edge node[left] {} (1);
    \path [-](2) edge node[left] {} (5);
    \path [-](4) edge node[left] {} (6);
    \path [-](1) edge node[left] {} (2);
    \path [-](4) edge node[left] {} (2);
    \path [-] (3) edge node[left] {} (4);
    \path [-] (3) edge node[left] {} (7);
    \path [-] (8) edge node[left] {} (4);
    \path [-] (6) edge node[left] {} (5);
    \path [-] (6) edge node[left] {} (9);
    \path [-] (8) edge node[left] {} (7);
    \path [-] (8) edge node[left] {} (9);
    \path [-] (10) edge node[left] {} (9);
    \path [-] (10) edge node[left] {} (11);
    \path [-] (11) edge node[left] {} (12);
    \path [-] (6) edge node[left] {} (13);
    \path [-] (13) edge node[left] {} (14);
    \path [-] (14) edge node[left] {} (15);
    \path [-] (10) edge node[left] {} (13);
    \path [-] (11) edge node[left] {} (14);
    \path [-] (12) edge node[left] {} (15);
    \path [-] (15) edge node[left] {} (18);
    \path [-] (14) edge node[left] {} (17);
    \path [-] (13) edge node[left] {} (16);
    \path [-] (5) edge node[left] {} (16);
    \path [-] (17) edge node[left] {} (18);
    \path [-] (16) edge node[left] {} (17);

\end{tikzpicture}
\caption{}
\label{fig:rectangle}
\end{subfigure}
%================================
\begin{subfigure}[b]{0.4\textwidth}
\centering 
\begin{tikzpicture}[scale=0.65]
    \node[shape=circle,draw=black, inner sep=1pt, outer sep = 0pt] (3) at (0,0) {3};
    \node[shape=circle,draw=black,draw=black, inner sep=1pt, outer sep = 0pt] (1) at (0,1.5) {1};
    \node[shape=circle,draw=black,draw=black, inner sep=1pt, outer sep = 0pt] (4) at (1.5,0) {4};
    \node[shape=circle,draw=black, inner sep=1pt, outer sep = 0pt] (2) at (1.5,1.5) {2};
    \node[shape=circle,draw=black,draw=black, inner sep=1pt, outer sep = 0pt] (5) at (3,1.5) {5};
    \node[shape=circle,draw=black, inner sep=1pt, outer sep = 0pt] (6) at (3,0) {6};
    \node[shape=circle,draw=black, inner sep=1pt, outer sep = 0pt] (8) at (1.5,-1.5) {8};
    \node[shape=circle,draw=black,draw=black, inner sep=1pt, outer sep = 0pt] (9) at (3,-1.5) {9};
    \node[shape=circle,draw=black,draw=black, inner sep=1pt, outer sep = 0pt] (7) at (0,-1.5) {7};

    \path [-] (4) edge node[left] {} (1);
    \path [-](4) edge node[left] {}  (5);
    \path [-](4) edge node[left] {}  (6);
    \path [-](9) edge node[left] {}  (4);
    \path [-](4) edge node[left] {}  (2);
    \path [-] (3) edge node[left] {} (4);
    \path [-] (4) edge node[left] {} (7);
    \path [-] (8) edge node[left] {} (4);
    %\path [-] (6) edge node[left] {} (5);
    %\path [-] (6) edge node[left] {} (9);
    %\path [-] (8) edge node[left] {} (7);
   % \path [-] (8) edge node[left] {} (9);

\end{tikzpicture}
\caption{}
\label{fig:star_graph}
\end{subfigure}

\caption{(a) star graph (b) 2D Regular Lattice }
\label{fig:graphs}
\end{figure}
%=========================================
\begin{figure}[!ht]
	\centering
	\begin{subfigure}{.4\textwidth}
		\includegraphics[width=7cm]{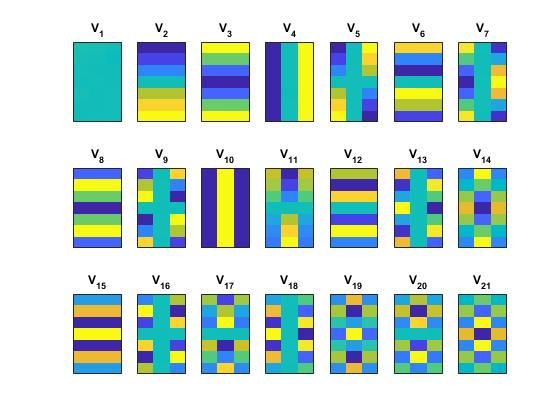}
		\caption{}
	\end{subfigure}
%%%%%%%%%%%%%%
	\begin{subfigure}{.4\textwidth}
		\includegraphics[width=7cm]{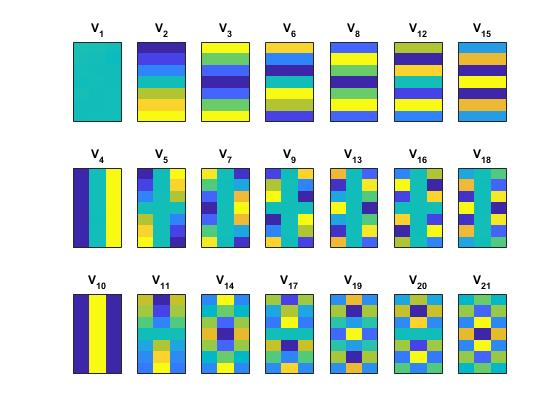}
		\caption{}
	\end{subfigure}
%%%%%%%%%%%%%%
\caption{(a) Eigenvectors of  rectangle grid graph sequentially ordered
in terms of nondecreasing eigenvalues.\\ (b) The ordering of the eigenvectors has been done from the paper \cite{saito2018can}.}
        \label{fig:Ordering_Grid}
        \end{figure}

%==============================
%============

%===========================

%===========================

%==================
%\subsection{Duality Motivation}

%We can attempt to establish a certain symmetry between graph-time and graph-frequency domains - in particular. We can require that graph Fourier transform components themselves must reside on vertices of a (frequency domain) graph. Such symmetry between time and frequency domains is captured by the concept of duality in traditional signal processing. The duality between time and frequency domains is an important property of Fourier transforms and has been heavily used in discrete-time analysis \cite{oppenheim1975digital}. Duality leads to the concept of Fourier transform pairs and helps in applying algorithms developed for the time domain to the frequency domain. Given a graph $\G$, we could potentially identify its dual $\G_d$. Ideally, for signals residing on the graph $\G$; their GFT resides on the dual graph $\G_d$, and applying GFT again gives us back the original signal on $\G$. A dual graph $\G_d$ describes the relations among the frequency components. Also dual graph solves the problem of ordering of the eigenvectors (mentioned earlier).

\section{Preliminaries and Notation}
%\begin{enumerate}
%\item  Definition of GSP, highlight non-uniqueness
%\end{enumerate}
%============================

Given an unweighted undirected graph $\G$, $A_\G$ denotes its adjacency matrix. Note that the adjacency matrix is symmetric with zero diagonal and non negative off diagonal entries. There exists a diagonal matrix (of eigenvalues) $\Lambda$ and an orthogonal matrix $V$ of eigenvectors such Adjacency or Laplacian matrix $A_\G$ or $L_\G = V \Lambda V^T.$ %$= \sum_{k}\lambda_k v_k v_k^\textsf{T}$. Here $v_1, v_2, \ldots, v_n$ represent the columns of $V$ and $\lambda_1, \lambda_2, \ldots, \lambda_n$ represent the diagonal entries of $\Lambda$. 
Given a graph signal $\vec{x}$ on $\G$, its GFT can be defined by $V^\Tr x$. Note in particular, as highlighted by previous works in \cite{huang2018graph} \cite{teke2017uncertainty} for example, that the GFT is not uniquely defined: for instance multiplying the columns of $V$ arbitrarily with $1$ and $-1$ also gives a valid eigenvector matrix. 

We denote by $\text{tr}(A)$ the trace of a square matrix $A$, and by $\norm{A}_F = \text{tr}(A^\He A)$ the Frobenius norm of a matrix $A$. Note that if $V, U $ are any unitary matrix ($V^\He V = I$, $U^\He U = I$) then $\norm{AV}_F = \norm{UA}_F = \norm{A}_F.$ A complex unit is a complex number with absolute value $1$. For a complex number $z$, the real part of $z$ is denoted by $\Re z$. For a matrix $A$, the entries are denoted by $A_{ij}$, the conjugate matrix by $\bar{A}$, and the matrix containing absolute values of the entries in $A$ by $|A|$. The Hadamard or element-wise product of matrices $A$ and $B$ is written $A \odot B$. We denote by $\bm{1}$ a vector of all $1'$s, by $\text{diag}(x)$ a diagonal matrix with $x$ as the diagonal, and by $\text{diag}(D)$ the diagonal of a matrix $D$.

Most of the simulations are carried out on both  Erd\H{o}s-R\'{e}nyi graphs ( we denote by $G(N,p)$ an Erd\H{o}s-R\'{e}nyi graph on $N$ nodes with edge probability $p$) and structured graphs (grid  and star graphs; see in Figure \ref{fig:graphs}).

\begin{figure}[!ht]
%\hspace{-2cm}
\includegraphics[width=8cm]{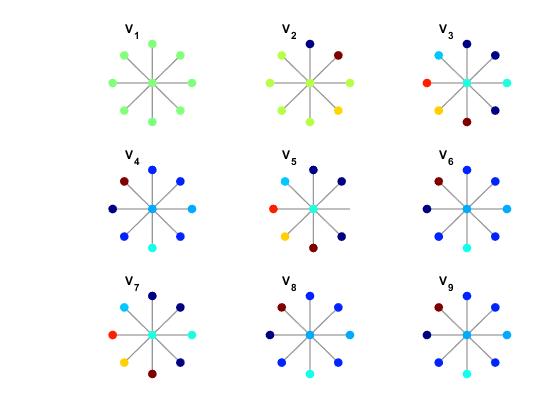}
\centering
\caption{ Eigenvectors of  star graph sequentially ordered
in terms of non-decreasing eigenvalues.}
\label{fig:seq_orderL}
\end{figure}

\vspace{-1cm}
\section{Prior work}
\subsection{Ordering graph eigenvectors}
The ordering on frequency indices can be determined by checking the similarity of the corresponding eigenvectors. For e.g. motivated by heat diffusion, the authors in \cite{cloninger2021dual} measure this similarity by taking the global average of local correlations. In equation \eqref{eq:diffusion} $\alpha(v_1,v_2)$ represents the similarity between vectors $v_1$ and $v_2$. Here $p(t,n,m)$ is classical heat kernel.  This pairwise similarity matrix (distance matrix) between frequency indices allows us to construct a graph in the frequency domain.

\begin{equation}
\label{eq:diffusion}
\alpha(v_1,v_2)^2=\norm{v_1v_2}_2^2 \sum_{n=1}^{N} ( \sum_{m=1}^{N} p(t,n,m)(v_1(m)-v_1(n))(v_2(m)-v_2(n)))^2.
\end{equation}

In \cite{saito2018can}, the authors construct the distance matrix based on Ramified optimal transport theory. Each eigenvector is converted to a probability mass function (pmf) in the first step. The distance is then computed by transporting one pmf measure to another pmf along ramified transport paths with some specific transport cost functional. Once the distance matrix is constructed, it is embedded into a low dimensional space to organize the eigenvectors properly. %The author investigated the ordering of the eigenvectors of the $2-D$ rectangular grid graph.  

%\begin{enumerate}
%\item  diffusion metric
%\item transport theory metric embedding to obtain the graph
%\end{enumerate}

\subsection{Duality in GSP}
\label{sec:prior-work-duality}
%\begin{enumerate}

%\item take transpose and obtain new graph. Issues of non-uniqueness.

%\item Compare ordering obtained by structure based techniques to ordering obtained by signal processing duality.

%\item (ignore ismorphism and other properties)
%\end{enumerate}

As discussed earlier, the signal processing dual of a graph $\G_1$ is a graph $\G_2$ such that a GFT on $\G_2$ inverts a GFT on $\G_1$. Let us first start with the following preliminary analysis by consider a graph signal on $\G_1$, and its GFT $V^\Tr x$ (assume without loss of generality that $V$ is real). This frequency domain signal now resides on $\G_2$, so applying the GFT again leads to $W^\Tr V^\Tr x$ (here $W$ is an eigenvector matrix of $\G_2$): so that to get back the original signal we require that $W = V^\Tr$. Thus an eigenvector matrix of the $\G_2$ must be the transpose of an eigenvector matrix of $\G_1$. We must construct $\G_2$ such that $V^\textsf{T}$ is an eigenvector matrix of $\G_2$. The eigenvalues of $\G_2$ are unknown, so they can be determined by solving an appropriate feasibility problem as in \cite{leus2021dual}. Let $\tilde{v}_k$ represent the $k^{\textsf{th}}$ column of $V$, then the adjacency matrix of the graph with eigenvector matrix $V^\textsf{T}$ is given by $A(\Lambda) = V^\textsf{T} \Lambda V = \sum_{k}\lambda_k \tilde{v}_k \tilde{v}_k^\textsf{T}$ where $\Lambda = \text{diag}(\lambda_1, \lambda_2, \ldots, \lambda_n)$ is the (yet unknown) matrix of eigenvalues of $\G_2$. We can find the unknown eigenvalues by solving the linear program

\begin{equation}
\label{eq:feasibility}
\begin{split}\Lambda = \argmin\{1 | A(\Lambda)_{ii}= 0, A(\Lambda) \geq 0, A(\Lambda)\bm{1} \geq \bm{1} \},
\end{split}
\end{equation}

Where the last constraint assumes no isolated nodes in the graph and imposes a normalization to avoid all zero solutions, similar solutions to find graphs with a given set of eigenvectors have been used in the graph learning literature \cite{dong2019learning}.

At this point, we would like to note that there is no guarantee that the linear program above is feasible. Indeed, with our experiments on ER graphs, we observe that for almost $1000$ graphs in $G(5, 0.5)$ (out of a total of $1024$ graphs), the feasibility problem above does not have a solution in $88\%$ of the cases. We will discuss more on this in the next section.

%================================
%\input{graph-fig}

\begin{figure}[htbp]
\centering 
 % \begin{tabular}{ccc}
\begin{subfigure}[b]{0.15\textwidth}
\centering
\begin{tikzpicture}[scale=0.4]
    \node[shape=circle,draw=black, inner sep=1pt, outer sep = 0pt] (3) at (0,0) {3};
    \node[shape=circle,draw=black, inner sep=1pt, outer sep = 0pt] (1) at (0,1.5) {1};
    \node[shape=circle,draw=black, inner sep=1pt, outer sep = 0pt] (4) at (1.5,0) {4};
    \node[shape=circle,draw=black, inner sep=1pt, outer sep = 0pt] (2) at (1.5,1.5) {2};
    \node[shape=circle,draw=black, inner sep=1pt, outer sep = 0pt] (5) at (1.5/2,-1.5) {5};

    \path [-] (3) edge node[left] {} (1);
    \path [-](3) edge node[left] {} (5);
    \path [-](4) edge node[left] {} (5);
    \path [-](1) edge node[left] {} (2);
    \path [-](4) edge node[left] {} (2);
    \path [-] (3) edge node[left] {} (4);
\end{tikzpicture}
\label{fig:graphs-dual1}
\caption{Graph  $\G_1$}
\end{subfigure}
%& 
\begin{subfigure}[b]{0.15\textwidth}
\centering
\begin{tikzpicture}[scale=0.4]
    \node[shape=circle,draw=black, inner sep=1pt, outer sep = 0pt] (1) at (0,0) {1};
    \node[shape=circle,draw=black, inner sep=1pt, outer sep = 0pt] (4) at (1.5,0) {4};
    \node[shape=circle,draw=black, inner sep=1pt, outer sep = 0pt] (3) at (3,0) {3};
    \node[shape=circle,draw=black, inner sep=1pt, outer sep = 0pt] (2) at (1.5,1.5) {2};
    \node[shape=circle,draw=black, inner sep=1pt, outer sep = 0pt] (5) at (1.5,-1.5) {5};

    \path [-] (4) edge node[left] {} (1);
    \path [-](3) edge node[left] {} (5);
    \path [-](1) edge node[left] {} (5);
    \path [-](1) edge node[left] {} (2);
    \path [-](3) edge node[left] {} (2);
    \path [-] (3) edge node[left] {} (4);
\end{tikzpicture}
\label{fig:graphs-dual2}
\caption{Graph  $\G_2$}
\end{subfigure}
%\tdplotsetmaincoords{65}{110}
\begin{subfigure}[b]{0.15\textwidth}
\centering
\begin{tikzpicture}[scale=0.4]
    \node[shape=circle,draw=black, inner sep=1pt, outer sep = 0pt] (2) at (0,0) {2};
    \node[shape=circle,draw=black, inner sep=1pt, outer sep = 0pt] (4) at (0,1.5) {4};
    \node[shape=circle,draw=black, inner sep=1pt, outer sep = 0pt] (5) at (1.5,0) {5};
    \node[shape=circle,draw=black, inner sep=1pt, outer sep = 0pt] (3) at (1.5,1.5) {3};
    \node[shape=circle,draw=black, inner sep=1pt, outer sep = 0pt] (1) at (1.5/2,-1.5) {1};

    \path [-] (2) edge node[left] {} (5);
    \path [-](4) edge node[left] {} (5);
    \path [-](4) edge node[left] {} (2);
    \path [-](4) edge node[left] {} (3);
    \path [-](3) edge node[left] {} (5);
    \path [-] (1) edge node[left] {} (2);
    \path [-] (1) edge node[left] {} (5);
\end{tikzpicture}
\label{fig:graphs-dual3}
\caption{Graph  $\G_3$}
\end{subfigure}
\caption{For the graph $\G_1$, the graph $\G_3$ is a more likely candidate for signal processing dual than graph $\G_2$, since $1.728 = \Du(\G_1, \G_2) > \Du(\G_1, \G_3) = 0.454$.}
\label{fig:graphs-dual}
\end{figure}

%================================
\section{Proposed dualness measure}
\label{sec:dualness_measure}
This section, we propose a measure of how close two graphs $\G_1$ and $\G_2$ are to being signal processing duals. If $V_1$ and $V_2$ are eigenvector matrices of $\G_1$ and $\G_2$ respectively, based on the discussion in Section \ref{sec:prior-work-duality}, we may start with $d(V_1, V_2) = \norm{V_1 - V_2^\Tr}_F = \norm{V_1V_2-I}_F$ as a potential dualness measure. However, this formulation has two key issues:
\begin{enumerate}
    \item A graph matrix $A_{\G}$ can have many possible eigenvector matrices. The choice of the eigenvector matrix could change the measure $d(V_1, V_2)$ above. Ideally, we expect the proposed measure to be invariant to the choice of the eigenvector matrix. See Figure \ref{fig:graphs-dual} as an example of how much the value of  $d(V_1, V_2)$ can vary depending on the choice of $V_1$ and $V_2$. 
    \item Suppose the vertices of $\G_1$ are ordered differently (i.e the rows and columns of the graph matrix  are permuted, then the eigenvector matrix would change, and so would the measure $d(V_1, V_2)$ above. We would like the proposed measure to be a graph invariant.
\end{enumerate}

Note that (1) can be seen to be the key reason behind the infeasibility of the linear program defined in \eqref{eq:feasibility}: a different choice of the eigenvector matrix may potentially provide a solution to the feasibility problem in \eqref{eq:feasibility}.

To formalize this, we let $\V_\G$ be the set of \emph{all} eigenvector matrices of a graph $\G$ (including various choices of the eigenvectors and labelings). We can then consider
\begin{equation*}
    %\label{eq:dualness-def}
    \min_{\substack{V_1 \in \V_{\G_1} \\ V_2 \in \V_{\G_2}}}\norm{V_1V_2-I}^2_F =  2-2\max_{\substack{V_1 \in \V_{\G_1} \\ V_2 \in \V_{\G_2}}}\Re\ \Tra(V_1V_2),
\end{equation*}
leading to the definition

\begin{equation}
   % \label{eq:dualness_trace}
      \label{eq:dualness-def}
    \Du(\G_1, \G_2) = \max_{\substack{V_1 \in \V_{\G_1} \\ V_2 \in \V_{\G_2}}}\Re\ \Tra(V_1V_2).
\end{equation}
We refer to $\Du(\G_1, \G_2)$ as the dualness measure between graphs $\G_1$ and $\G_2$. As discussed above, this is invariant to labelings and the choice of eigenvectors by design. The higher the value of $\Du(\G_1, \G_2)$, the closer the graphs are to being signal processing duals. Note that since $V_1$ and $V_2$ are orthogonal, the largest possible value of $\Du(\G_1, \G_2)$ is $N$ (the number of nodes in the graph), and this largest value is attained when $V_1$ and $V_2$ are inverses of each other. In this case, $\G_1$ and $\G_2$ are signal processing duals.
%$\V_\G$ contains many matrices.
%and the dual $\G^*$ of a graph $\G$ may be defined as
%\begin{equation}
 %   \label{eq:dual-def}
 %   \G^* = \argmin_{\mathcal{H}}\Du(\G, \mathcal{H}).
%\end{equation}

%Note that the definition in \eqref{eq:dual-def} is potentially ambiguous, since there could be many graphs achieving the smallest dualness (this ambiguity can be partly resolved by Proposition \ref{prop:dual-iso} below). Thus this definition identifies many possible graphs as duals for a given graph (in particular the dual may not be unique, and $\G^*$ may be seen as the set of all possible duals).

For arbitrary graphs, finding $\Du(\G_1, \G_2)$ from  \eqref{eq:dualness-def} is discrete and non-convex in nature. We focus on finding algorithms to approximate $\Du(\G_1, \G_2)$, and this is the topic for the rest of this section.

%\section{Proposed dualness measure}
%===========================================
%\begin{enumerate}
 %    \item  Same as before. Example on why this complicated formulation is required (why not product of V). Need the measure to not change even if graph vertices are enumerated differently, or even if the eigenvectors are calculated differently.

%\item We give two algorithms to approximate the metric above. The second algorithm is more accurate, at the expense of more computations.
%\end{enumerate}

%==============================================
\subsection{Algorithm 1 (CD)}

\label{sec:alt-max-noperm}
We start with solutions to $\Du(\G_1, \G_2)$ under some assumptions, which though sub-optimal, helps in laying the groundwork for solutions in the general scenario.
 
We first observe that if all the eigenvalues of $A_\G$ are distinct, then all the eigenspaces are of dimension $1$. Given an eigenvector matrix $V \in \V_\G$, multiplying the columns of $V$ with any unit complex number and randomly permuting the columns gives all the matrices in $\V_\G$. Thus we first start with an assumption that the eigenvalues of $\G_1$ and $\G_2$ are all distinct and let $\D$ be the set of all diagonal matrices with complex units along the diagonal. Then we have
\begin{equation}
\label{eq:dualness-lb2}
    \Du(\G_1, \G_2)  \geq \max_{\substack{D_1, D_2 \in \D \\ \text{permutations }P_1, P_2}} \Re\ \Tra\left(V_1D_1P_1V_2D_2P_2\right).
\end{equation}

For a graph with distinct eigenvalues, the above inequality is tight.

First, let us ignore the permutations: We will later build on the discussion below to include the permutations in Section \ref{sec:matchpairs}. We then get a looser lower bound
\begin{equation}
\label{eq:dualness-lb1}
   \Du(\G_1, \G_2) \geq \max_{D_1, D_2 \in D} \Re\ \Tra\left(V_1D_1V_2D_2\right).
\end{equation}
%Note that due to the orthogonality of the matrices involved, the optimum of \eqref{eq:dualness-trace-form} is upper bounded by $N$. We will work with this max trace form of the objective of \eqref{eq:dualness-trace-form} in the subsequent sections.
We find the lower bound in \eqref{eq:dualness-lb1} using a coordinate descent algorithm. Coordinate descent or alternate maximization algorithms are a popular paradigm for solving optimization problems with many sets of variables \cite{wright2015coordinate}. Applied to \eqref{eq:dualness-lb1}, we start with random initialization of $D_1, D_2$ and iteratively try to pick $D_1$ and $D_2$ that maximize the objective. Suppose $D_1$ is fixed, then objective in \eqref{eq:dualness-lb1} is of the form $\Re\ \Tra(A D_2)$ where $A= V_1D_1V_2$. Given $A$, the choice of $D_2$ that maximizes the objective is given by $D_2 = \text{diag}\left(\overline{A}\odot (1/|A|)\right)$: the value of the objective with this choice is $\sum |A_{ii}|$. Similary if $D_2$ is known, the objective is  $\Re\ \Tra (V_1D_1V_2D_2)  = \Re\ \Tra (D_1V_2D_2V_1)$ by the cyclic property of trace. This is summarized in Algorithm \ref{alg:altmax}.

As discussed earlier, the objective is bounded above by $N$. Since each step of the algorithm only increases the objective, convergence is not an issue; however, the rate of convergence might be slow, and the limit may depend on the initialization. See Section \ref{sec:experiments} for some simulation results.

%\subsection{Max cut based algorithm}
%\cite{goemans1995improved}
\subsection{Algorithm 2 (CDPM)}
\label{sec:matchpairs}
In this section, we build on the development of Algorithm 1 (CD) to give a more general algorithm to compute  Consider the lower bound to dualness in \eqref{eq:dualness-lb2}. We wish to compute
\begin{equation}
    \label{eq:dualness-trace-permutation-form}
    \max_{\substack{D_1, D_2 \in D \\ \text{permutations }P_1, P_2 }} \Re\ \Tra\left(V_1D_1P_1V_2D_2P_2\right).
\end{equation}
The matrices $D_1P_1$ and $D_2P_2$ have exactly one non zero entry in each row and column, and all the non zero entries are all complex units. We can apply a technique similar to Algorithm \ref{alg:altmax}, where we alternately optimize over $D_1, P_1$ and $D_2, P_2$. Suppose $D_1$ and $P_1$ are known, the objective in \eqref{eq:dualness-trace-permutation-form} is of the form $\Re\ \Tra\left(AD_2P_2\right) = \Re\ \Tra\left(P_2AD_2\right)$. If we pick a permutation matrix $P_2$ corresponding to the permutation $\sigma$, i.e. $(P_2)_{i \sigma(i)} = 1$, then the value of $D_2$ that maximises the objective should be selected in the same way as before: we set $D_2 = \text{diag}\left(\overline{P_2A}\odot (1/|P_2A|)\right)$, and the value of the objective for this choice is \(\sum |A_{\sigma(i) i}|\). We now try to pick the permutation $\sigma$ such that this summation is maximised. Note that selecting the optimum $\sigma$ is equivalent to picking exactly one entry from each row and each column of $|A|$ such that the sum of all these entries is maximised. This is the well studied linear assignment problem \cite{burkard2009assignment}, which can be solved by polynomial time algorithms \cite{munkres1957algorithms}. This procedure can be repeated to find the optimum $P_1, D_1$ for a given $P_2, D_2$.  

Based on this, we can construct an alternate maximization algorithm similar to CD, summarized as Algorithm \ref{alg:altmax2}.

%==========================================
\begin{minipage}{0.46\textwidth}
\begin{algorithm}[H]
    \centering
    \caption{Coordinate descent based algorithm (CD)}
    \label{alg:altmax}
    \begin{algorithmic}
    \STATE \textbf{Given:} Graphs $G_1, G_2$ with $N$ nodes, eigenvector matrices 
    \STATE \textbf{Initialisation:} Pick $D_1,D_2 \in \D$ randomly 
    \STATE Pick eigenvector matrices $V_1, V_2$ of $\G_1$ and $\G_2$
    %\STATE $\textbf{Pseudo Code:}$
    %\STATE $[V_1,e_1] \leftarrow eig(G_1)$
    %\STATE $[V_2,e_2] \leftarrow eig(G_2)$
    \STATE $\text{currentObj} = \Re\ \Tra(V_1D_1V_2D_2), \text{previousObj} = -\infty$ 
    \WHILE{$ \mid{\text{currentObj} - \text{previousObj}}\mid \leq \epsilon$}
    %\STATE
    %\STATE $\COMMENT{\text{for finding }D_2}$
    %\STATE $S_1 \leftarrow V_1D_1V_2$ 
    \STATE $D_2\leftarrow \text{diag}\left(\overline{V_1D_1V_2}\odot (1/|V_1D_1V_2|)\right)$
   % \STATE
    %\STATE $\COMMENT{\text{for finding }D_1}$
    %\STATE $S_2 \leftarrow V_2D_2V_1$
    
    \STATE $D_1\leftarrow \text{diag}\left(\overline{V_2D_2V_1}\odot (1/|V_2D_2V_1|)\right)$
    
    \STATE $\text{previousObj} \leftarrow \text{currentObj}$
    \STATE $\text{currentObj} \leftarrow \Re\ \Tra(V_1D_1V_2D_2)$
    \ENDWHILE
    \end{algorithmic}
    \end{algorithm}
\end{minipage}
\hfill
\begin{minipage}{0.49\textwidth}
\begin{algorithm}[H]
    \centering
    \caption{Coordinate descent and perfect matching based algorithm (CDPM)}
\begin{algorithmic}
 \label{alg:altmax2}
\STATE Given graphs $G_1$ and $G_2$ each with $N$ nodes
\STATE \textbf{Initialisation:} Pick $D_1, D_2 \in \D$, permutation matrices $P_1, P_2$ randomly
%\STATE $\textbf{Pseudo Code:}$
\STATE Pick eigenvector matrices $V_1, V_2$ of $\G_1$ and $\G_2$
\STATE $\text{currentObj} \leftarrow \Re \Tra(V_1D_1P_1V_2D_2P_2)$,
\STATE $\text{previousObj} \leftarrow -\infty$,
\WHILE{$ \mid{\text{currentObj} - \text{previousObj}}\mid \leq \epsilon$}
\STATE $S_2 \leftarrow V_1D_1P_1V_2$ 
\STATE Solve the assignment problem on $|S_2|^\Tr$ to obtain $P_2$
\STATE $D_2\leftarrow \text{diag}\left(\overline{P_2S_2}\odot (1/|P_2S_2|)\right)$
\STATE $S_1 \leftarrow V_2D_2P_2V_1$
\STATE Solve the assignment problem on $|S_1|^\Tr$ to obtain $P_1$
\STATE $D_1\leftarrow \text{diag}\left(\overline{P_1S_1}\odot (1/|P_1S_1|)\right)$
\STATE $\text{previousObj} \leftarrow \text{currentObj}$
\STATE $\text{currentObj} \leftarrow \Re \Tra(V_1D_1P_1V_2D_2P_2)$
\ENDWHILE
\end{algorithmic}
\end{algorithm}
\end{minipage}

%========================================

\begin{comment}

 \begin{algorithm}
\caption{Coordinate descent and perfect matching based algorithm (CDPM)}
\begin{algorithmic}
 \label{alg:altmax2}
\STATE Given graphs $G_1$ and $G_2$ each with $N$ nodes
\STATE \textbf{Initialisation:} Pick $D_1, D_2 \in \D$, permutation matrices $P_1, P_2$ randomly
%\STATE $\textbf{Pseudo Code:}$
\STATE Pick eigen-vector matrices $V_1, V_2$ of $\G_1$ and $\G_2$
\STATE $\text{currentObj} \leftarrow \Re \Tra(V_1D_1P_1V_2D_2P_2)$,
\STATE $\text{previousObj} \leftarrow -\infty$,
\WHILE{$ \mid{\text{currentObj} - \text{previousObj}}\mid \leq \epsilon$}
\STATE $S_2 \leftarrow V_1D_1P_1V_2$ 
\STATE Solve the assignment problem on $|S_2|^\Tr$ to obtain $P_2$
\STATE $D_2\leftarrow \text{diag}\left(\overline{P_2S_2}\odot (1/|P_2S_2|)\right)$
\STATE $S_1 \leftarrow V_2D_2P_2V_1$
\STATE Solve the assignment problem on $|S_1|^\Tr$ to obtain $P_1$
\STATE $D_1\leftarrow \text{diag}\left(\overline{P_1S_1}\odot (1/|P_1S_1|)\right)$
\STATE $\text{previousObj} \leftarrow \text{currentObj}$
\STATE $\text{currentObj} \leftarrow \Re \Tra(V_1D_1P_1V_2D_2P_2)$
%\STATE
\ENDWHILE
\end{algorithmic}
\end{algorithm}

\end{comment}

\subsection{Validation}
\label{sec:experiments}

\begin{figure}[!ht]
	\centering
	\begin{subfigure}{.4\textwidth}
		\includegraphics[width=7cm]{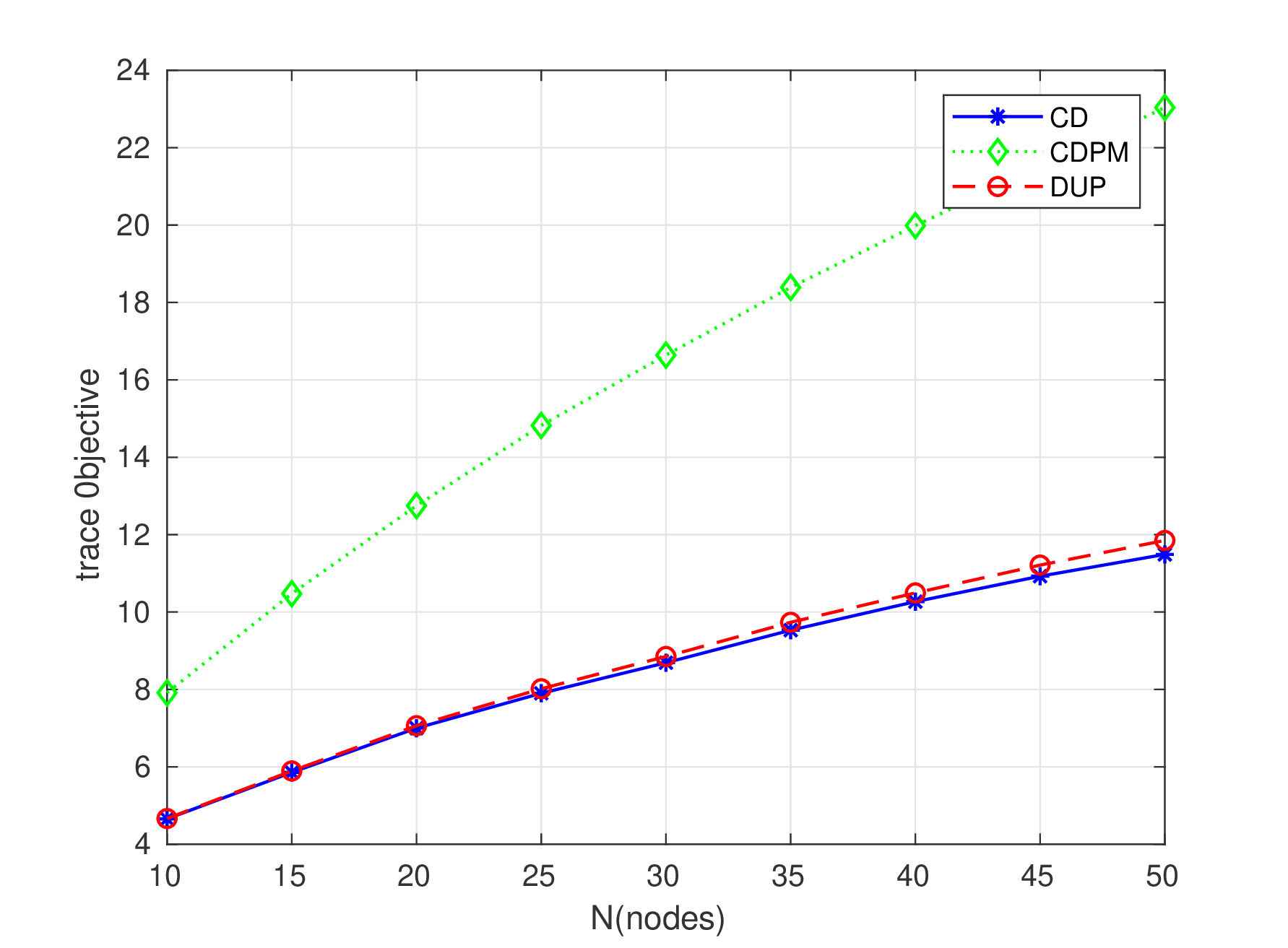}
		\caption{}
		\label{fig:figure1}
	\end{subfigure}
%%%%%%%%%%%%%%
	\begin{subfigure}{.4\textwidth}
		\includegraphics[width=7cm]{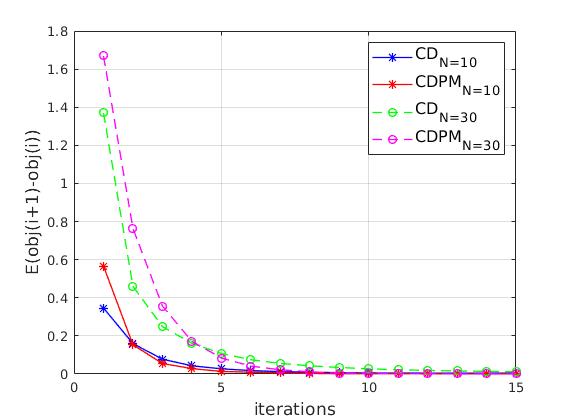}
		\caption{}
		\label{fig:figureiter}
	\end{subfigure}
%%%%%%%%%%%%%%
\caption{(a)Comparison of CD (Algorithm \ref{alg:altmax}) and CDPM (Algorithm \ref{alg:altmax2} for graphs in $G(N, 0.4)$. The upper bound (DUP) for the optimization problem  in \eqref{eq:dualness-lb1} is also plotted. (b)Comparison of Expected rate of convergence of CD Algorithm $1$) and CDPM (Algorithm $2$) over iterations for
graphs in $G(10, 0.4)$ and $G(30, 0.4)$.}
        \label{}
        \end{figure}

%Check with the duality bound.
%=================================
In this section, we validates our algorithms to compute the approximate values of $\Du(\G_1,\G_2)$ by comparing the obtained values with a duality-based upper bound. We generate a pair of Erd\H{o}s-R\'{e}nyi (ER) graphs in $G(N,p)$ with distinct eigenvalues and apply the algorithms above to compute the approximate dualness of this pair. This exercise is repeated about $100$ times, and the average is computed. A plot of this average trace objective vs. $N$ is shown for both the algorithms in Figure \ref{fig:figure1}.

To evaluate the performance of CD (Algorithm \ref{alg:altmax}), we compute an upper bound on the objective in \eqref{eq:dualness-lb1} by evaluating its dual. If we consider the diagonal entries of $D_1, D_2$ as variables, we see that the objective is a quadratic form and can be expressed as $x^\He W x$ for 
\[
x = \begin{pmatrix}\text{diag}(D_1) \\ \text{diag}(D_2) \end{pmatrix}_{2N \times 1} \text{ and } W = \frac{1}{2}\begin{pmatrix}0 & V_1^\Tr \odot V_2\\ V_1 \odot V_2^\Tr&  0\end{pmatrix}.
\]
The problem in \eqref{eq:dualness-lb1} is thus equivalent to maximizing $x^\He W x$ when the elements of $x$ are restricted to be complex units. When $x$ is forced to be real, the entries of $x$ can only be $\pm 1$, reducing this problem to the max cut problem \cite{bollobas2013modern}. The dual of this problem can be checked to be an SDP \cite{boyd2004convex}, solving which gives us an upper bound to the optimal value in \eqref{eq:dualness-lb1}.
%\begin{equation}
%\label{eq:dual-bnd}
%\mathfrak{u}(\G_1, \G_2) = \min \left\{1^\Tr\nu\ | -W + %\text{diag}(\nu) \geq 0\right\} \text{(DUP)}.
%\end{equation}
From Fig \ref{fig:figure1}, we observe the plot for CD follows the plot for DUP very closely - the coordinate descent based algorithm performs near-optimally for values of $N$ from $10$ to $50$.

%=========================================================

Figure \ref{fig:figureiter} shows the convergence of proposed algorithms (the difference between successive iterations, averaged over initial conditions and problem input). We observe that both algorithms converge fairly fast. For e.g. with $N=30$, we observe the algorithms converge within $15$ (for CD) and $10$ (for CDPM) iterations for most initializations. However, the limit point does depend on the initialization, so the algorithm does get stuck on local maxima. To evaluate the performance in Fig \ref{fig:figure1}, the best solution among $200$ initializations was picked.

%==========================

%\section{Evaluating CDPM}
%\label{sec:evaluating-cdpm}
%In Figure \ref{fig:figure1} we compared the performance of the Algorithm \ref{alg:altmax} (CD) to the theoretical upper bound by computing the dual of \eqref{eq:dualness-trace-form}. A similar approach can be adopted to evaluate the dual of the generalized trace objective in \eqref{eq:dualness-trace-permutation-form}. By treating the entries of the permutation matrices as the variable $q$, and with a suitable rearranging of dimensions, the objective in \eqref{eq:dualness-trace-permutation-form} can be written as an optimization problem with quadratic objective and constraints:
%\begin{mini}
%	{q}{-q^\He W q}{\label{opt:1}}{}
%		\addConstraint{\sum_{i \in I}|q_i|^2 }{=1 \text{ for indices }I\text{ in the same row}}
 %   	\addConstraint{q_{i}q_{j}}{=0 \ \text{ for }q_i, q_j \text{ in the same row}.}
%\end{mini}
 
%The dual of this optimization program can be written as an SDP; however it seems that the dual optimum is always $N$ (which is the maximum possible value of the primal objective). It stands to reason that the duality gap for \eqref{eq:dualness-trace-permutation-form} itself seems to be high, and better techniques may be required to test the optimality of CDPM. Also, the dual SDP scales very poorly with $N$, as it has  $O(n^3)$ variables.

\section{Analysis and observations}
\label{sec:Analysis}

In this section, we evaluate the dualness measure (using Algorithm 2) for the dual graphs generated by distance-based methods \cite{saito2018can}, \cite{cloninger2021dual}, and the graphs generated by signal processing based methods \cite{leus2021dual}. We start with ER graphs with $N$ nodes (in the range $10$ to $40$)  and $p=0.3$. Starting with this graph as $\G_1$, the dual graph $\G_2$ for different methods is computed. The distance-based techniques from \cite{saito2018can}, \cite{cloninger2021dual} provide a distance metric between the frequency indices, which we convert to graph edge weights by applying a gaussian kernel function $exp(-d^2/2\sigma^2)$ where $\sigma=0.5$. The dualness measure $\Du(\G_1,\G_2)$ is computed (using Algorithm $2$) and averaged over $200$ graphs in $\G(N,p)$. Table \ref{tab:table1} evaluates the dualness measure based on Algorithm $2$ (CDPM). We see that the signal processing-based methods \cite{leus2021dual} to obtain the dual graph results in a higher dualness measure compared to the distance-based techniques. For reference, we also compare with an algorithm that returns a random graph in $G(N,p)$ as dual. This is along the expected lines since our dualness measure is designed for signal processing-based techniques.

However, when we evaluate the proposed algorithms for structured graphs, a rectangular grid graph with $N=21$ nodes and a star graph with $N=9$ nodes, we notice a somewhat opposite trend. The Table \ref{tab:table2} shows the signal processing dual method Leus \cite{leus2021dual} has less dualness measure compared to distance-based techniques \cite{saito2018can}, \cite{cloninger2021dual}. This suggests that for structured graphs, there is scope to improve the existing methods to find the frequency domain graph, possibly by using ideas from the distance-based methods.

%======================================
\begin{minipage}[c]{0.5\textwidth}
\centering
\begin{tabular}{|c|c|c|c|c|}
\hline
 \multicolumn{5}{|c|}{Dualness measures Comparison on CDPM} \\
\hline
Nodes  & 10 & 20 & 30 & 40 \\ 
Cloninger \cite{cloninger2021dual} & 6.49 & 12.21 & 16.30 & 20.12 \\
Saito \cite{saito2018can}  & 7.26 & 12.47 & 16.43  & 20.40\\
Leus \cite{leus2021dual}  & 8.54 & 13.85 & 17.64 & 21.45\\
Random graph  & 6.28 & 12.32 & 16.25 & 20.15\\
\hline

\end{tabular}
\captionof{table}{ Dualness measure for the ER graph}
\label{tab:table1}
\end{minipage}
\begin{minipage}[c]{0.5\textwidth}
\centering
\begin{tabular}{|c|c|c|}
\hline
\multicolumn{3}{|c|}{Dualness measures Comparison on CDPM} \\
\hline
Graph & Star & Grid \\
Nodes & 9 & 21\\
Cloninger  \cite{cloninger2021dual} & 6.21 & 12.85 \\
Saito \cite{saito2018can} & 8.546 &  13.78 \\
Leus \cite{leus2021dual} & 6.79 & 12.96 \\
\hline
\end{tabular}
\captionof{table}{Dualness measure for the structured graphs.}
\label{tab:table2}
\end{minipage}

%==========================

\section{Conclusion and future work}

We proposed algorithms to compare different graphs/ordering on the frequency indices. The proposed dualness measure checks how close two graphs are to having inverse GFTs. This measure is invariant to node relabelling and the choice of the eigenvector matrix. By comparing with existing techniques to evaluate frequency domain graphs, we observed that there is scope to use distance/similarity-based techniques to improve GSP-based duality techniques.

The algorithms in Section \ref{sec:dualness_measure} were developed assuming all the eigenvalues of $\G_1$ and $\G_2$ are distinct. However, many structured graphs do have repeated eigenvalues, including for e.g., a complete graph, clustered graph, star graph, ring graph, etc \cite{chung1997spectral}. In this case, the eigenspaces have more than one orthogonal vector, and consequently, the space $\V_\G$ contains a lot more matrices. We hope to generalize Algorithm \ref{alg:altmax2} to include repeated eigenvalues and obtain a tighter bound to the dualness measure.

 Other problems to work on include techniques to test the optimality of CDPM, evaluate the performance of both CD and CDPM algorithms for large values of $N$, find the dual from dualness measures, and proofs of optimality. It might be interesting to find some interpretations for the dual graph and subsequent applications to graph signal processing.
 \vspace{-0.8cm}
 %The grant number for this project is ECR/2018/001832.
\section{Acknowledgment}
We would like to thank the SCIENCE $\&$ ENGINEERING RESEARCH BOARD (SERB), India, for the financial support to this work.

\end{document}